# Evolution of self-assembled InAs/Gas(001) quantum dots grown by growth-interrupted molecular beam epitaxy

**Adalberto Balzarotti**[‡]

Dipartimento di Fisica, Università di Roma "Tor Vergata", Via della Ricerca Scientifica, I-00133 Roma, Italy.

[‡] E-mail: balza@roma2.infn.it

**Abstract**

Self-assembled InAs quantum dots (QDs) grown on GaAs(001) surface by molecular beam epitaxy under continuous and growth-interruption modes exhibit two families of QDs, quasi-3D (Q3D) and 3D QDs, whose volume density evolution is quantitatively described by a rate-equation kinetic model. The volume density of small Q3D QDs decreases exponentially with time during the interruption, while the single-dot mean volume of the large QDs increases by Ostwald ripening. The kinetics of growth involves conversion of quasi-3D to 3D QDs at a rate determined by superstress and participation of the wetting layer adatoms. The data analysis excludes that quasi-3D QDs are extrinsic surface features due to inefficient cooling after growth.

**1.Introduction**

InAs/GaAs is a prototype of the strained Stranski-Krastanov epitaxial growth in which the stress energy is relaxed by the nucleation of quantum dots (QDs) of different density and volume[1]. At the typical growth temperature of 500°C, a couple-of-monolayers (ML) high islands - sometimes termed quasi-3D quantum dots (Q3D QDs) - start nucleating on the wetting layer (WL) at deposition of InAs around 1.45 ML and sit preferentially at the upper edges of 2D islands. With increasing coverage, a second family of dots (3D QDs) grows on terraces and undergoes an explosive nucleation with an order of magnitude change in density within 0.2 ML above a critical coverage $\theta_c \sim 1.57$ ML. Their volume is larger than the deposited volume because of a sizable surface-mass transport during growth. The WL is also involved and its thickness is reduced progressively by the supply of adatoms to QDs until the barrier for nucleation increases and the nucleation process ends [2]. We have previously used periodic growth interruption (GI) of the In flux to monitor the surface-mass transport [3], the size distribution of QDs [4], their scale invariance [5], and the

WL erosion around QDs nucleated on steps [6]. While the bimodal distribution of dots is now accepted [7,8], the nature of the quasi-3D islands is still debated [9] and a quantitative description of the nucleation and growth processes which would enable the precise prediction of their evolution is lacking or limited to the subcritical ($\theta<\theta_c$) region [10]. On the other hand, the understanding of the exact mechanism of self-organization of these nanostructures is crucial for the development of QDs-based nanoelectronic and optoelectronic devices [11]. Therefore we analyze the volume density of InAs QDs on GaAs(001) substrates obtained by scanning atomic force microscopy (AFM) with a growth model based on kinetic rate equations and apply it to continuous (CG) and GI growth modes to establish the influence of the in situ annealing without deposition flux on the volume density and size of QDs.

## 2. Experimental details

InAs films of thickness ranging between 1.45 and 2.2 ML were grown by molecular-beam epitaxy (MBE) on non-rotating GaAs(001) substrates by taking advantage from the angular dependence of the flux to achieve coverage differences as small as 0.01 ML. The samples were grown at 500 °C with a growth rate of 0.029 MLs$^{-1}$ by cycling the In delivery in 5 s of evaporation followed by 25 s of GI with the As shutter left open. Many topographs of area ranging between (0.3x0.3) and (5x5) μm², were examined by ex-situ AFM in tapping mode to measure the density and volume of QDs as a function of coverage. Further experimental details are reported elsewhere [2]. In Figure1 an AFM image of the surface shows the various features and their respective line profile [12].

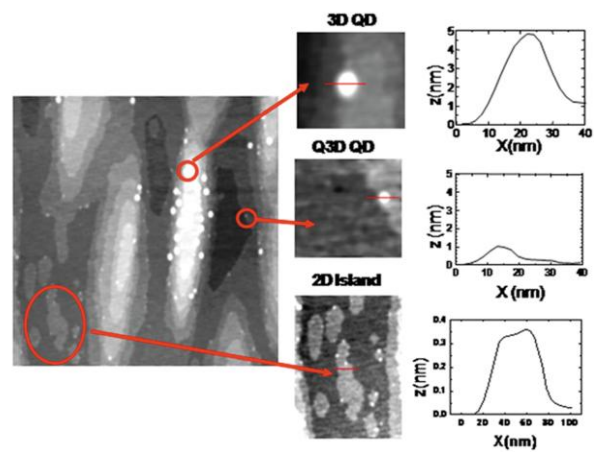

**Figure1**.(Colour online) (1x1) μm² AFM image of the GaAs(001) surface for an InAs coverage of 1.6 ML. Molecular beam epitaxy with growth interruption at 500 °C has been used. Various structural features are evidenced together with the corresponding line profiles measured along the indicated lines. The large 2D clusters on the terraces are 1-ML(0.28 nm)-thick and they correspond to an as-yet incomplete wetting layer. The mounds have average length along $[1\bar{1}0]$ of 1.3 nm, width along [110] of 830 nm and height of 120 nm.

## 3. Results and discussion

*3.1. Continuous growth mode volume density*

As an example of CG, we first consider the experiment of Ramachandran et al. [7] who measured the morphology of the nanostructures in the range $\theta$=1.35-2.18 ML crossing $\theta_c$~1.57 ML. The authors identify two dimensional (2D) platelets, Q3D islands (2-4 ML high) and 3D dots at higher coverage and suggest that Q3D islands could act as precursors to the 3D QDs.

In order to describe the time evolution of the amounts of adatoms, $n_1$, precursors, $n_2$, and quantum dots, $n_3$, on the surface, we write the kinetic rate equations in the mean-field approximation [10,13]:

$$\begin{cases} \dot{n}_1 = F - \beta(n_1 + n_3) - n_1(\kappa_2 n_2 + \kappa_3 n_3) \\ \dot{n}_2 = \kappa_2 n_1 n_2 - \gamma n_2 \\ \dot{n}_3 = \beta n_3 + \kappa_3 n_1 n_3 + \gamma n_2 \,, \end{cases} \quad (1)$$

where F is the InAs flux (ML s$^{-1}$), $\kappa_j=\sigma_j$ D (j=2,3) are the rate coefficients for adatom attachment to precursors and QDs, respectively, $\sigma_j$ are the size-normalized diffusion capture factors and D is the surface diffusion coefficient of adatoms. The detachment rate of adatoms from precursors is small and can be disregarded, whereas the direct capture rate of adatoms, $\beta$, by the WL and QDs is unity at 500 °C [14]. In models of strained epitaxy based on the classical theory of nucleation [15,16], the mass transfer is a thermally activated process described by the nucleation rate, $\gamma$, which, in turn, depends exponentially on the superstress $\zeta=(\theta-\theta_c)/\theta_c$. $\zeta$ measures the instability (2D to 3D transition) of the WL for thickness larger then the equilibrium thickness [17]. Recently, Song et al. [10] used $\gamma=\gamma_0 \exp[\zeta E/kT]$, where $\gamma_0$ and E are constant parameters to fit the high-energy-electron-diffraction intensity from the WL of InAs deposited on GaAs for $\theta<\theta_c$. They found $\gamma_0$=0.09±0.02 s$^{-1}$ and E=2.0±0.3 eV and a direct dependence of E on the deposition flux at small growth rates. E$\zeta$ represents the formation energy of QDs.

In Figure2, we compare the volumes (in unit of ML) of the InAs nanostructures on GaAs(001) substrates as a function of $\theta$, calculated by solving the system of Eqs.(1), with the corresponding experimental distributions [7]. As initial conditions we assume that adatoms are dynamically in equilibrium with the WL under the flux F before the appearance of precursors and QDs and that precursors, in absence of an exact criterion to fix their critical size [13], have the measured density [7] lower than 4x10$^{-7}$nm$^{-2}$ around $\theta$=1.4 ML.

Using the optimized set of parameters given by Song et al.[10], we find good agreement between our calculations and the experimental data of Ramachandran et al.[7]. It is worth noting that the model supports strongly the binomial size distribution of self-assembled QDs of InAs on GaAs.

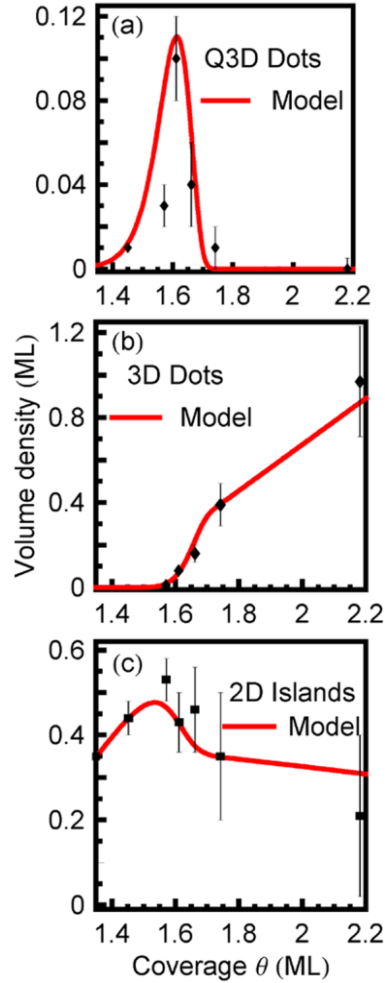

**Figure2**.(Colour online) Coverage dependence of the volume density (in ML) of InAs (a) quasi-3D quantum dots $n_2$, (b) 3D quantum dots $n_3$, and (c) 2D islands $\theta-1-(n_2+n_3)$, growing over a 1-ML wetting layer of GaAs(001). Scattered symbols with error bars are the experimental data from Ref.[7] measured at 500 °C using a growth rate F= 0.22±0.022 ML s$^{-1}$ and continuous growth (CG) mode. Solid curves are calculated according to the model (1). The parameters values are E=3eV, $n_1(t_0)$=0.22 ML, $n_2(t_0)$= 6x10$^{-4}$ ML, $n_3(t_0)$=0, $t_0$=5.6 s, and the size-normalized capture rates $\kappa_2$=20, $\kappa_3$=40.

In detail, the transition from the 2D-island morphology of the WL to the 3D morphology of QDs occurs in the narrow range 1.45-1.74 ML in which the WL increases consuming the adatoms and the precursors grow up to θ~1.61 ML. Below $\theta_c$, ζ is negative and the metastable WL [18] continues to grow layer-by-layer under the InAs flux. The relieve of the strain energy takes place through the progressive increase of the height of the 2D islands to form Q3D clusters with small aspect ratio. Beyond $\theta_c$, ζ turns to positive and the nucleation rate rises exponentially with θ starting from the value γ ≈ 0.09 s$^{-1}$. At θ= 1.61 ML the superstress is large enough that nucleation of 3D QDs with higher aspect ratio starts dominating because taller clusters are more relaxed than flatter ones [19]. Concurrently, the precursors drop abruptly down to 1.74 ML and convert to QDs at high rate. We find that at θ∼2.2 ML the contribution from the impinging flux after the 2D-3D transition to the final QD volume of 0.89 ML is 0.69 ML and the rest, 0.1 ML comes from precursors and 0.1 ML from the adatoms directly attaching to QDs. Therefore part of the mass transfer to QDs comes from the precursors in agreement with Ref.[7]. For θ≥1.8 ML the decrease of the 2D island volume with increasing InAs delivery signals the attachment (at a rate β) of adatoms to QDs (Figure2(c)). As remarked by Ramachandran et al.[7] this is consistent with the disappearance beyond 1.74 ML of the large 2D islands in the WL as well as of precursors, as evident from Figure2(a). Moreover, postgrowth ripening of the QDs must be excluded since the surface morphology was frozen very quickly after growth [14].

*3.2. Interrupted growth mode volume density*

The above quantitative analysis enables us to understand the main modifications which occur during the GI process in our experiment. Once the In flux is stopped after 5 s of deposition, precursors are allowed to convert into QDs without refilling from the adatoms. The rate equation for $n_2$ can approximately be written as

$$\dot{n}_2 = -\gamma n_2, \qquad (2)$$

because the adatom volume density $n_1$ is small at the early stages of growth and γ is nearly constant and equal to $\gamma_0$ for F=0. Thus the volume of Q3D dots decreases exponentially with time with a time constant $\tau=\gamma^{-1}$ during the 25 s of growth interruption. The amount of InAs contained in the small Q3D clusters, calculated from Eq. (2), taking as initial values for $n_2$ at $t_0$=5 s the ones obtained by solving the system (1), is shown in figure 3(a). It is found that the Q3D volume density is reduced by more than an order of magnitude compared to that of CG, in close agreement with the experimental data.

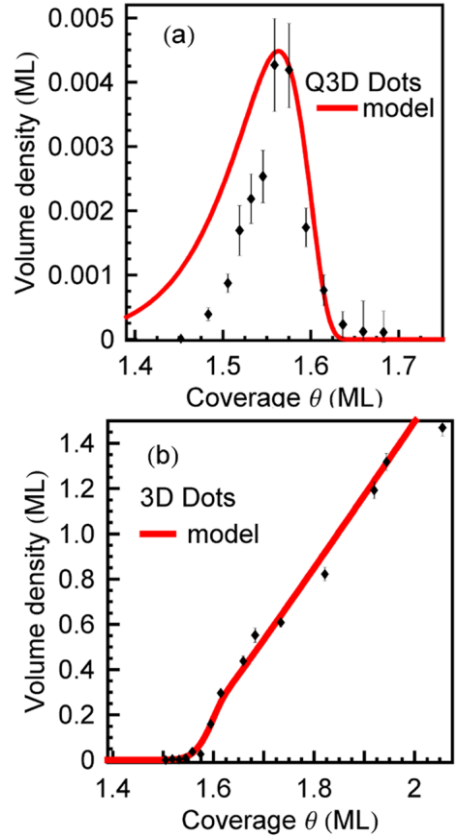

**Figure 3**.(Colour online) (a): Coverage dependence of the volume density (in ML) of InAs/ GaAs(001) quasi-3D quantum dots calculated from Eq.(2) after 25 s of growth interruption and (b): 3D quantum dots volumes $n_3$ scaled by a constant factor of ~2. The scattered symbols with error bars are the experimental data measured at $500^0$C using a growth rate F=0.029 MLs$^{-1}$ and interrupted growth (GI) mode. The parameters values are E=6eV, $n_1(t_0)$=0.029 ML, $n_2(t_0)$=6x10$^{-4}$ ML, $n_3(t_0)$=0, $t_0$=45 s, and the size-normalized capture rates $\kappa_2$=20, $\kappa_3$ =40.

Above $\theta_c$ the calculated $n_3$ values must be scaled up by a factor of about 2 to agree with the experiment, as shown in Figure 3(b). This result points to the Ostwald ripening of the 3D QD volume during the 25 s nominal anneal time at zero flux much larger than in CG, alike that observed after 30 s anneal at 485 °C [9]. The average volumes of the dots measured as a function of coverage are presented in Figure 4(a). The 3D QD volume increases quickly above $\theta_c$ and saturates around 880 nm$^3$ for $\theta$>1.8 ML. Moreover, for $\theta$=1.58 ML we find [20] that the surface, after prolonged annealing at 500 °C in vacuum, has the speckled appearance shown in Figure4(b), where the circled area highlights truncated dots of approximate 0.2 nm height. It is likely that they are the remnant of the precursors which have lost material to form 3D QDs [21].

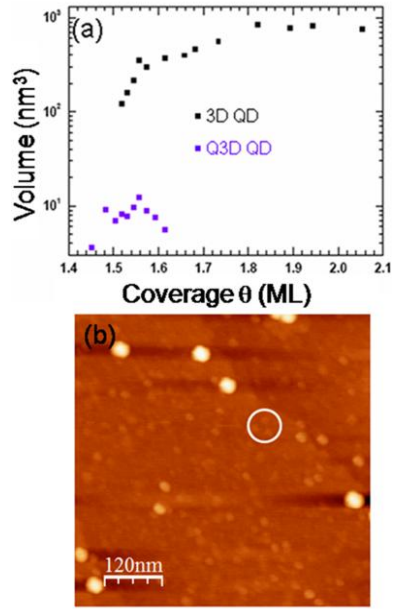

**Figure 4**.(Colour online) (a): Measured average volume V (in nm$^3$) of the quasi-3D and 3D quantum dots as a function of InAs coverage θ. Small Q3D quantum dots have an average height 0.2 nm ≤h ≤1 nm and width w around 20 nm, while large quantum dots have h≥2 nm and w≤20 nm. (b): 0.6x0.6 μm$^2$ AFM image for θ=1.58 ML after 30 min in situ annealing at 500°C; the circled area highlights islands of height h ≤0.2 nm.

The volume of these latter dots is 860 nm$^3$ and their density is an order of magnitude lower than that of the unannealed sample, indicating that the volume saturates provided enough time is given to adatoms to diffuse. This fact emphasizes the important rôle played by surface diffusion in the ripening of QDs. The fast ripening of the 3D QDs is concurrent with the decrease of the 2D island volume (not shown here), indicating that the WL adatoms contribute considerably to the kinetics until the coalescence of QDs sets up at coverages of ~2 ML, to which the model calculation deviates from the experiment, as apparent in Figure3(b). We stress that the above analysis excludes that the bimodal size distribution of QDs is an extrinsic effect of adatom condensation due to postgrowth quenching [9, 22] even though a minor contribution to their final size cannot be entirely ruled out. Compared with CG, the annealing process during GI is responsible for both the exponential decay of precursors and the ripening of 3D QDs.

## 4. Conclusions

To summarize, we have shown that the Stranski-Krastanov growth kinetics of coherent self-assembled QDs in InAs/GaAs(001) is strongly influenced by the parameters of growth. In the GI mode the average volume density of quasi-3D dots decays exponentially with time during flux interruption with strain-dependent rate, leaving a number of dots much smaller compared to CG growth. Moreover, the Ostwald ripening of the 3D QDs is stronger on account of the large surface mass transfer from the WL. Quantitative agreement with experiment is found for the volume density of quantum dots and 2D islands using a mean-field model of growth and a set of tested parameters. The conversion of quasi-3D to QDs reinforces the strain-relieving mechanism of migration of edge atoms to the top of the 2D islands to form precursors [13] and supports the intrinsic nature of these nanostructures.

**Acknowledgment**

This work was supported by the COFIN 2005 project No. 2005025173 of the Italian Ministry of Research.

References

[1] For a review, see: Joyce B.A., Vvedensky D.M. *Materials Science and Engineering* 2004 R **46**, 127

[2] Placidi E., Arciprete F., Fanfoni M., Patella F., and Balzarotti A. 2007 *J.Phys.Condens.Matter* **19**, 225006

[3] Patella F., Arciprete F., Fanfoni M., Sessi V., Balzarotti A., Placidi E. 2005 *Appl.Phys.Lett.* **87**, 252101

[4] Patella F., Nufris S., Arciprete F., Fanfoni M., Placidi E., Sgarlata A., and Balzarotti A. 2003 *Phys.Rev. B* **67**, 205308

[5] Fanfoni M., Placidi E., Arciprete F., Orsini E., Patella F., and Balzarotti A. 2007 *Phys.Rev.B* **75**, 205312

[6] Placidi E., Arciprete F., Sessi V., Fanfoni M., Patella F., and Balzarotti A. 2005 *Appl. Phys. Lett.* **86**, 241913

[7] Ramachandran T.R., Heitz R., Chen P., and Madhukar A. 1997 *Appl.Phys.Lett.* **70**, 640

[8] Patella F., Arciprete F., Placidi E., Nufris S., Fanfoni M., Sgarlata A., Schiumarini D., and Balzarotti A. 2002 *Appl.Phys.Lett.* **81**, 2270

[9] Kryzewski T.S. and Jones T.S. 2004 *J.Appl.Phys.* **96**, 668

[10] Song H.Z., Usuki T., Nakata Y., Yokoyama N., Sasakura H., and Muto S. 2006 *Phys.Rev.B* **73**, 115327

[11] Schmid G.(Ed.) 2004 *Nanoparticles: From Theory to Applications* Wiley-VCH, Weinheim

[12] On surfaces with a small density of steps prepared by heating the sample at 650 °C for 50 min under $As_4$ flux of $2 \times 10^{-5}$ Torr [see LaBella V.P. *et al*. 2000 *Phys.Rev.Lett.* **84**, 4152 ] and subsequently depositing an InAs buffer layer, we find that the quasi-3D volume density is unaffected and the 3D QD volume density is slightly reduced above $\theta > 1.8$ ML.

[13] Dobbs H.T., Vedensky D., Zangwill A., Johansson J., Carlsson N., and Seifert W. 1997 *Phys.Rev.Lett.* **79**, 897

[14] Kobayashi N.P., Ramachandran T.R, Chen P., and Madhukar A. 1996 *Appl.Phys.Lett.* **68**, 3299

[15] Osipov A.V., Schmitt F., Kukushkin S.A., Hess P. 2002 *Appl.Surf.Sci.* **188**, 156

[16] Dubrovskii V.G., Cirlin G.E., and Ustinov V.M. 2003 *Phys.Rev.B* **68**, 075409

[17] Muller P. and Kern R. 1996 *Appl.Surf. Sci*. **102**, 6

[18] Eisenberg H.R., and Kandel D. 2000 *Phys.Rev.Lett.* **85**, 1286

[19] Shchukin V.A., Ledentsov N.N., Kop'ev P.S., and Bimberg D. 1995 *Phys.Rev.Lett.* **75**, 2968

[20] Della Pia A. 2008 Undegraduate thesis, University of Roma "Tor Vergata", unpublished

[21] The average aspect ratio decreases from 0.25 to 0.15 relative to the unannealed surface case. It is defined as $h/w = 2.3\, \pi^{1/2}\, V/S^{3/2}$, where V and S area the volume and area of the dots, respectively, and 2.3 is an experimental constant value intermediate between a spherical cap (2) and a cone (3) or truncated pyramid. Nevertheless, details of island shape are immaterial since they do not significantly alter our results.

[22] Tsukamoto S., Honma T., and Arakawa Y. 2006 *Small* **3**, 386